\documentclass{aastex}
\usepackage{spr-astr-addons}
\usepackage{url}
\usepackage{amsfonts}
\usepackage{amssymb,epsfig}
\usepackage{latexsym}
\usepackage{amsmath}
\usepackage{graphicx}

\RequirePackage{color}

\begin{document}

\title{Cosmological evolution and statefinder diagnostic for new holographic dark energy model in non flat universe}

\slugcomment{Not to appear in Nonlearned J., 45.}
\shorttitle{Short article title} \shortauthors{Autors et al.}

\author{M. Malekjani\altaffilmark{1,2}} \and \author{A. Khodam-Mohammadi\altaffilmark{1}} \and \author{N. Nazari-pooya\altaffilmark{1}}
\altaffiltext{}{E-mail: malekjani@basu.ac.ir.}
\altaffiltext{}{E-mail: khodam@basu.ac.ir.} \altaffiltext{}{E-mail:
Nazarip@basu.ac.ir}
 \altaffiltext{1}{Department of Physics, Faculty
of Science, Bu-Ali Sina University, Hamedan 65178, Iran.}

\altaffiltext{2}{Research Institute for Astronomy $\&$ Astrophysics
of Maragha (RIAAM), Maragha, Iran.}


\begin{abstract}
In this paper, the holographic dark energy model with new infrared
cut-off proposed by Granda and Oliveros
 has been investigated in spatially non flat universe. The dependency of the evolution of equation
of state, deceleration parameter and cosmological evolution of
Hubble parameter on the parameters of new HDE model are calculated.
Also, the statefinder parameters $r$ and $s$ in this model are
derived and the evolutionary trajectories in $s-r$ plane are
plotted. We show that the evolutionary trajectories are
 dependent on the model parameters of new HDE model. Eventually, in the light of
SNe+BAO+OHD+CMB observational data, we plot the evolutionary
trajectories in $s-r$ and $q-r$ planes for best fit values of the
parameters of new HDE model.
\end{abstract}

\section{Introduction\label{intro}}
The observational evidences from distant Ia supernova, Large Scale
Structure (LSS) and Cosmic Microwave Background (CMB) suggest that
our universe is undergoing an accelerating expansion \citep{SN}.
Within the framework of general relativity this expansion may be
driven by a component with negative pressure, namely, dark energy
(DE) \citep{copel, fri}. The nature of DE is still unknown and many
theoretical models have been provided to describe
 the cosmic behavior of DE. The simplest and important theoretical
candidate for DE is the Einstein's cosmological constant, which can
fit the observations well so far \citep{berg89,sahnicc}. The
cosmological constant suffers two well known problems namely
"fine-tuning" and "cosmic coincidence" \citep{copel}.  In order to
alleviate or even solve these problems, many dynamical dark energy
models have been proposed, whose equation of state is time-varying.
Dynamical DE models can be classified into two categories i) The
scalar field dark energy models including quintessence
\citep{Wetterich}, K-essence \citep{Chiba}, phantoms
\citep{Caldwell1}, tachyon \citep{Sen}, dilaton \citep{Gasperini},
quintom \citep{Elizalde1} and so forth. ii) The interacting dark
energy models, by considering the interaction between dark matter
and dark energy, including Chaplygin gas \citep{Kamenshchik},
braneworld models \citep{Deffayet}, holographic DE (HDE)
\citep{holo1} and agegraphic
DE  (ADE)\citep{age1} models and so forth.\\
 On the other hand, a curvature driven acceleration model which is called, modified gravity, has
 been proposed by Strobinsky \citep{star} and Kerner \citep{kerner} et al., for the first time, in 1980. Modified
 gravity approach suggests the gravitational alternative for unified description of inflation,
 dark energy and dark matter with no need of the hand insertion of extra dark
 components \citep{setare}.\\
 The HDE model is constructed based on the holographic
principle \citep{Cohen,Horava,holo1,Li}. In holographic principle,
based on the validity of the effective local quantum field theory, a
short distance (UV) cut-off $\Lambda $ is related to the long
distance (IR) cut-off $L$ due to the limit set by the formation of a
black hole \citep{Cohen}. The HDE model has been constrained by
various astronomical observation
\citep{obs3a,obs2,obs1,Wu:2007fs,obs3} and also investigated widely
in the literature \citep{nonflat,holoext,intde}. The HDE model with
Hubble horizon or particle horizon as a length scale, can not derive
the accelerated expansion of the universe \citep{holo1}. Although,
in the case of event horizon, HDE model can derive the universe with
accelerated expansion \citep{Li}, but the arising problem with the
event horizon is that it is a global concept of spacetime and
existence of it depends on the future evolution of the universe only
for a universe with forever accelerated expansion. Moreover, the HDE
with the event horizon is not compatible with the age of some old
high redshift objects \citep{zhang4}. The above problems with HDE
motivated us to follow the new HDE model proposed by Granda and
Oliveros (GO, here after). GO proposed a new IR cut-off containing
the local quantities of Hubble and time derivative Hubble scales
\citep{garanda}. The advantages of HDE with GO cutoff (new HDE, here
after) is that it depends on local quantities and avoids the
causality problem appearing with event horizon IR cutoff. The new
HDE model can also obtain the accelerated expansion of the universe
\citep{garanda}. GO showed that in new HDE model, the transition
redshift from deceleration phase ($q>0$) to acceleration phase
($q<0$) is
consistent with current observational data \citep{garanda,observ}.\\
Besides, the observational experiments such as CMB experiment
\citep{Sie} and luminosity - distance of supernova measurements
\citep{Caldwell} imply that our universe is not perfectly flat and
has a small positive curvature. Therefore, we
are motivated to consider the new HDE in the non-flat universe.\\
Since many dynamical DE models have been proposed to interpret the
cosmic acceleration, a sensitive and diagnostic tool is required to
discriminate the various DE models. The various DE models have a
degeneracy on the Hubble parameter $H$ (first time derivative of
scale factor) and the deceleration parameter $q$ (second time
derivative of scale factor). Therefore, these quantities cannot
discriminate the DE models. For this aim, we need the higher order
of the time derivative of scale factor. By using the third order
time derivative, Sahni, et al. \citep{Sahni} and Alam et
al.\citep{alam} introduced the statefinder pair \{r,s\} in order to
remove the degeneracy of $H_0$ and $q_0$ of  different DE models.
The statefinder pair \{r,s\} is defined as
\begin{equation}
r=\frac{\dddot{a}}{aH^{3}},~~~~~~~s=\frac{r-\Omega_{tot}}{3(q-\Omega_{tot}/2)},
\label{rspair}
\end{equation}
where $\Omega_{tot}$ is the dimensionless total energy density containing matter
, DE and curvature.
It is clear that the statefinder is a geometrical diagnostic,
because it depends  on the scale factor. The role of statefinder
pair is to distinguish the behaviors of cosmological evolution of
dark energy models with the same values of $H_{0}$ and $q_{0}$ at
the present time.
 Up to now, the statefinder diagnostic tool has been used to study
the various dark energy models. The statefinder has been used to
diagnose different cases of the model, including different model
parameters and various spatial curvature contributions. The various
DE models have different evolutionary trajectories in \{r,s\} plane.
For example, the well-known $\Lambda$CDM model corresponds to a
fixed point $\{r=1,s=0\}$ in \{r,s\} plane \citep{Sahni}. Also, the
quintessence DE model \citep{Sahni,alam}, the interacting
quintessence models \citep{r12,r13}, the holographic dark energy
models
\citep{r14,r15}, the holographic dark energy model in non-flat universe \citep%
{r16}, the phantom model \citep{r18}, the tachyon \citep{r22}, the
agegraphic DE model with and without interaction in flat and
non-flat universe \citep{wei,malek} and the interacting new
agegraphic DE model in flat and non-flat universe
\citep{zhang,khod10}, are analyzed through the statefinder
diagnostic
tool.\\
 In this paper, we study the cosmological treatment of
 new HDE model and investigate this model by means of statefinder diagnostic. The paper is organized as
follows: In section 2, we present the new HDE model and derive the
statefinder parameters \{r,s\} for this model. In section 3, the
numerical results are presented. We conclude in section 4.

\section{New HDE model \label%
{theory}}
 Following \citep{garanda}, the energy density of new HDE is written
 as
 \begin{equation}
\rho_{\Lambda}=3M_{P}^{2}(\alpha H^{2}+\beta\dot{H}),\label{rhoh}
\end{equation}
where $\alpha$ and $\beta$ are constants, $M_p$ is the reduced Plank
mass, $H$ is the Hubble parameter and dot denotes the
derivative with respect to the cosmic time.\\
Here we assume the Friedmann-Robertson-Walker (FRW) universe as
follows:
\begin{eqnarray}
 ds^2=dt^2-a^2(t)\left(\frac{dr^2}{1-kr^2}+r^2d\Omega^2\right),\label{metric}
 \end{eqnarray}
 where $a(t)$ is the scale factor, and $k = -1, 0, 1$ represent the
open, flat, and closed universe, respectively. The observational
evidence reveal a closed universe with small positive curvature
($\Omega_k\sim 0.02$) \citep{Bennett}.  The first Freidmann equation
for a universe with curvature $k$ is written as
\begin{eqnarray}\label{Fried}
H^2+\frac{k}{a^2}=\frac{1}{3M_p^2}(\rho_m+\rho_{\Lambda}),
\end{eqnarray}
where $\rho_{\Lambda}$ is the energy density of DE and $\rho_m$ is
the energy density of matter including the cold dark matter ( CDM)
and baryons. Substituting the energy density of new HDE, i.e.
Eq.(\ref{rhoh}), in (\ref{Fried}) and changing the variable from the
cosmic time $t$ to $x=\ln{a}$, yields
\begin{equation}\label{HDE1}
H^2(1-\alpha)+\frac{k}{a^2}-\frac{\beta}{2}
\frac{dH^2}{dx}=\frac{\rho_{m0}}{3M_p^2}e^{-3x},
\end{equation}
where we assume the evolution of matter component as $\rho_m=\rho_{m0}e^{-3x}$.
 With the definition of normalized Hubble parameter as $E=H/H_0$, $\Omega_k=-k/H_0^2$,
 and $\Omega_{m0}=\rho_{m0}/3M_p^2H_0^2$, we
 can rewrite Eq. (\ref{HDE1}) as follows
 \begin{equation}
 E^2(1-\alpha)=\Omega_ke^{-2x}+\frac{\beta}{2}\frac{dE^2}{dx}+\Omega_{m0}e^{-3x}
\end{equation}
Solving the above first order differential equation for $E^2$ and
using the initial condition $E_0=1$, we obtain the dimensionless
Hubble parameter, $E$, for a universe containing the new HDE and
matter, as follows \citep{wang2}
\begin{eqnarray}\label{E1}
E^2=\Omega_ke^{-2x}+\Omega_{m0}e^{-3x}+\Omega_{\Lambda}(x),
\end{eqnarray}
where $\Omega_{\Lambda}(x)$ is the dimensionless energy density of
new HDE model which is given as  \citep{wang2}
\begin{eqnarray}\label{omeg_new}
&\Omega_{\Lambda}=\frac{\alpha-\beta}{-\alpha+\beta+1}\Omega_ke^{-2x}+\frac{2\alpha-3\beta}{-2\alpha+3\beta+2}\Omega_{m0}e^{-3x}\nonumber\\
&+(1-\frac{1}{-\alpha+\beta+1}\Omega_k-\frac{2}{-2\alpha+3\beta+2}\Omega_{m0})e^{-\frac{2(\alpha-1)x}{\beta}}
\end{eqnarray}
Inserting Eq. (\ref{omeg_new}) in (\ref{E1}), the parameter $E$ can
be obtained as
\begin{eqnarray}\label{solv}
&E=\Big(\frac{1}{-\alpha+\beta+1}\Omega_ke^{-2x}+\frac{2}{-2\alpha+3\beta+2}\Omega_{m0}e^{-3x}\nonumber\\
&+(1-\frac{1}{-\alpha+\beta+1}\Omega_k-\frac{2}{-2\alpha+3\beta+2}\Omega_{m0})e^{-\frac{2(\alpha-1)x}{\beta}}\Big)^{1/2}
\end{eqnarray}
Here we consider the parameters  $\beta\neq0$ and $\alpha\neq1$. The special cases when the denominators in
Eq. (\ref{solv}) are equal zero have been discussed in Ref. \citep{wang2}.\\
Combining Eq. (\ref{omeg_new}) with the conservation equation of new
HDE model, we can obtain the EoS parameter of new HDE as follows
\begin{eqnarray}\label{w_newhde}
&w_{\Lambda}=-1-\frac{1}{3}\frac{d\ln{\Omega_{\Lambda}}}{dx}=-1+\frac{2}{3}\times\\
\nonumber&
\Big[\frac{\alpha-\beta}{-\alpha+\beta+1}\Omega_ke^{-2x}+\frac{3}{2}\frac{2\alpha-3\beta}{-2\alpha+3\beta+2}\Omega_{m0}e^{-3x}+\\
\nonumber
&\frac{(\alpha-1)}{\beta}(1-\frac{1}{-\alpha+\beta+1}\Omega_k-\frac{2}{-2\alpha+3\beta+2}\Omega_{m0})e^{-\frac{2(\alpha-1)x}{\beta}}\Big]\\
\nonumber&
\Big{/}\Big[\frac{\alpha-\beta}{-\alpha+\beta+1}\Omega_ke^{-2x}+\frac{2\alpha-3\beta}{-2\alpha+3\beta+2}\Omega_{m0}e^{-3x}+\\
\nonumber
&(1-\frac{1}{-\alpha+\beta+1}\Omega_k-\frac{2}{-2\alpha+3\beta+2}\Omega_{m0})e^{-\frac{2(\alpha-1)x}{\beta}}\Big]
\end{eqnarray}
which is time-dependent EoS parameter. The time-dependent of EoS
parameter allows it to transit from $w_{\Lambda}>-1$ to
$w_{\Lambda}<-1$ \citep{wang22}. Some recent observational evidences
suggest DE models with whose EoS parameter crossees $-1$ in the near
past \citep{alam3}. In the limiting case of flat universe, by
considering the DE dominated epoch ( the matter contribution is
negligible compare with the contribution of DE), Eq.
(\ref{w_newhde}) is reduced as
\begin{equation}
w_{\Lambda}=-1+\frac{2(\alpha-1)}{3\beta}
\end{equation}
which is same as Eq. (2.5) in  \citep{granda2}.\\
Now we derive the deceleration parameter $q$ for new HDE model. The
deceleration parameter is given by
\begin{equation}
q=-\frac{\dot{H}}{H^2}-1
\end{equation}
Re-witting $q$ in terms of dimensionless Hubble parameter, $E$, we
have
\begin{equation}\label{q1}
q(x)=-\frac{1}{E}\frac{dE}{dx}-1
\end{equation}
Substituting  $E$ from Eq. (\ref{solv}) in (\ref{q1}), the parameter
$q$ can be obtain as
\begin{eqnarray}\label{qnew}
&q=
\Big[\frac{2}{-\alpha+\beta+1}\Omega_ke^{-2x}+\frac{3}{-2\alpha+3\beta+2}\Omega_{m0}e^{-3x}+\\
\nonumber &
(\alpha-1)(1-\frac{1}{-\alpha+\beta+1}\Omega_k-\frac{2}{-2\alpha+3\beta+2}\Omega_{m0})e^{-\frac{2(\alpha-1)x}{\beta}}\Big]\\
\nonumber &
\Big{/}\Big[\frac{1}{-\alpha+\beta+1}\Omega_ke^{-2x}+\frac{2}{-2\alpha+3\beta+2}\Omega_{m0}e^{-3x}+\\
\nonumber &
(1-\frac{1}{-\alpha+\beta+1}\Omega_k-\frac{2}{-2\alpha+3\beta+2}\Omega_{m0})e^{-\frac{2(\alpha-1)x}{\beta}}\Big]
\end{eqnarray}

Here, one can explicitly see the dependence of deceleration
parameter $q$ on the model parameters $\alpha$ and $\beta$.\\
Finally, we derive the statefinder pair \{r,s\} for new HDE model.
Using the definition of statfinder parameters in Eq. (\ref{rspair}),
we have
\begin{equation}
r=\frac{\dddot{a}}{aH^3}=\frac{\ddot{H}}{H^3}-3q-2.
\end{equation}
Similar to $q$, the parameter $r$ is re-written in terms of
dimensionless Hubble parameter, $E$, as
\begin{equation}\label{r1}
r=\frac{1}{E}\frac{d^2E}{dx^2}+\frac{1}{E^2}(\frac{dE}{dx})^2+\frac{3}{E}\frac{dE}{dx}+1,
\end{equation}
Using the definition of statefinder parameter $s$ in Eq.
(\ref{rspair}) and also $\Omega_{tot}=1+\Omega_k$, the parameter $s$
can be obtained in terms of $E$ as
\begin{equation}\label{s1}
s=-\frac{\frac{1}{E}\frac{d^2E}{dx^2}+\frac{1}{E^2}(\frac{dE}{dx})^2+\frac{3}{E}\frac{dE}{dx}+\Omega_k}{\frac{3}{E}\frac{dE}{dx}+\frac{3}{2}\Omega_k+\frac{9}{2}}
\end{equation}
Substituting $E$ from Eq. (\ref{solv}) in (\ref{r1}) and (\ref{s1}),
we obtain the parameters $r$ and $s$ for new HDE model

\begin{eqnarray}\label{r2}
&r=1+\frac{-2\alpha+3\beta+2}{\beta^2}\times  \\ \nonumber
&\Big[(\frac{1}{-\alpha+\beta+1}\Omega_k+\frac{2}{-2\alpha+3\beta+2}\Omega_{m0}-1)(\alpha-1)e^{-\frac{2(\alpha-1)x}{\beta}}\\
\nonumber&
-\frac{\beta^2}{(-\alpha+\beta+1)(-2\alpha+3\beta+2)}\Omega_ke^{-2x}\Big]\\
\nonumber&
\Big{/}\Big[-(\frac{1}{-\alpha+\beta+1}\Omega_k+\frac{2}{-2\alpha+3\beta+2}\Omega_{m0}-1)(\alpha-1)e^{-\frac{2(\alpha-1)x}{\beta}}\\
\nonumber &
+\frac{1}{-\alpha+\beta+1}\Omega_ke^{-2x}+\frac{2}{-2\alpha+3\beta+2}\Omega_{m0}e^{-3x}
\Big]
\end{eqnarray}

\begin{eqnarray}\label{s2}
&s=\frac{2}{3}\times\Big[(\frac{1}{-\alpha+\beta+1}\Omega_k+\frac{2}{-2\alpha+3\beta+2}\Omega_{m0}-1)\\
\nonumber&
(\frac{\alpha-1}{\beta}(-2\alpha+3\beta+2)+\beta\Omega_k)e^{-\frac{2(\alpha-1)x}{\beta}}\\
\nonumber&
-\beta \Omega_k (\frac{1+\Omega_k}{-\alpha+\beta+1}e^{-2x}+\frac{2}{-2\alpha+3\beta+2}\Omega_{m0}e^{-3x})\Big]\\
\nonumber& \Big{/}\Big[(\frac{1}{-\alpha+\beta+1}\Omega_k+\frac{2}{-2\alpha+3\beta+2}\Omega_{m0}-1)\\
\nonumber&
(-2\alpha+3\beta+2+\beta\Omega_k)e^{-\frac{2(\alpha-1)x}{\beta}}\\
\nonumber&
 -\beta \Omega_k(\frac{1+\Omega_k}{-\alpha+\beta+1}e^{-2x}+\frac{2}{-2\alpha+3\beta+2}\Omega_{m0}e^{-3x})
\Big]
\end{eqnarray}
Eqs. (\ref{r2},\ref{s2}) show the dependency of the statefinder pair
\{r,s\} on the model parameter of new HDE model $\alpha$ and $\beta$
as well as the curvature parameter $\Omega_k$.\\
Recently, Wang and Xu \citep{wang2}, by applying the Markov Chain
Monte Carlo method on the latest observational data, have
constrained the new HDE model. The observational data that have been
used are: the constitution dataset \citep{397Constitution} including
397 type supernova Ia (SNIa), the observational Hubble data (OHD)
\citep{OHD}, the cluster X-ray gas mass fraction
\citep{ref:07060033}, the measurement results of baryon acoustic
oscillation (BAO) from Sloan Digital Sky Survey (SDSS) \citep{SDSS}
and Two Degree Field Galaxy Redshift Survey (2dFGRS)
\citep{ref:Percival2}, and the cosmic microwave background (CMB)
data from five-year WMAP \citep{5yWMAP}. In non flat universe, the
best fit values of the new HDE model parameters ($\alpha$, $\beta$)
and the cosmological parameters ($\Omega_bh^2$, $H_0$,
$\Omega_k$,$\Omega_{\Lambda 0}$, $\Omega_{m0}$) with their
confidence level are obtained as:
$\alpha=0.8824^{+0.2180}_{-0.1163}$ ($1\sigma$)
$^{+0.2213}_{-0.1378}$ $(2\sigma)$,
$\beta=0.5016^{+0.0973}_{-0.0871}$ ($1\sigma$)
$^{+0.1247}_{-0.1102}$ $(2\sigma)$,
$\Omega_bh^2=0.0228^{+0.0010}_{-0.0010}$ ($1\sigma$)
$^{+0.0014}_{-0.0014}$ $(2\sigma)$, $H_0=70.20^{+3.03}_{-3.17}$
($1\sigma$) $^{+3.58}_{-4.00}$ $(2\sigma)$,
$\Omega_k=0.0305^{+0.0092}_{-0.0134}$ ($1\sigma$)
$^{+0.0140}_{-0.0176}$ $(2\sigma)$,
$\Omega_{\Lambda0}=0.6934^{+0.0364}_{-0.0304}$ ($1\sigma$)
$^{+0.0495}_{-0.0413}$ $(2\sigma)$ and
$\Omega_{m0}=0.2762^{+0.0278}_{-0.0320}$ ($1\sigma$)
$^{+0.0402}_{-0.0412}$ $(2\sigma)$.\\
In next section, we use the above best fit values of $\alpha$ and
$\beta$ for studying the cosmological behavior  of new HDE and also
the evolutionary trajectories of this model in $s-r$ plane.
\section{Numerical results\label{NR}}
In this section we give the numerically description of the
cosmological evolution and the statefinder trajectories in $s-r$
plane for new HDE model in spatially non flat universe. Here we use
the best fit values of the model parameters of new HDE as well as
the best fit values of cosmological parameters discussed in previous
section. The evolution of EoS parameter, deceleration parameter and
dimensionless Hubble parameter, $E$ of new HDE in non-flat universe
is calculated. Also, the evolutionary behavior of new HDE in $s-r$
plane is performed. Solving Eq. (\ref{w_newhde}), the evolution of
EoS parameter $w_{\Lambda}$ as a function of scale factor $a$ is
shown in Fig. (1). In both panels, we see that $w_{\Lambda}$ starts
from zero at the early time, representing the CDM dominated
universe, and crosses the phantom divide ($w_{\Lambda}<-1$) later.
In upper panel, by fixing $\alpha$  with the constrained
observational value $0.8824$, we vary the parameter $\beta$. The EoS
parameter $w_{\Lambda}$ becomes larger for smaller value of $\beta$.
We also see that the new HDE model crosses the phantom divide
earlier, for larger value of $\beta$. In lower panel, by fixing
$\beta$ with the constrained observational value $0.5016$, the
parameter $\alpha$ is varied. Unlike $\beta$, The EoS parameter
$w_{\Lambda}$ becomes larger, for larger value of $\alpha$. The
phantom divide is achieved earlier, for smaller value of $\alpha$. Here we showed the dependency
of the EoS of new HDE on the parameters of model.\\

In Fig.(2), using Eq.(\ref{qnew}), the evolution of deceleration
parameter $q$ as a function of scale factor $a$ is plotted. In upper
panel, we fixed $\alpha=0.8824$
 and varied the parameter $\beta$. The
transition from decelerated phase ($q>0$) to accelerated phase
($q<0$) takes place sooner, by increasing the parameter $\beta$.
Also, at any cosmic scale factor, the parameter $q$ is smaller by
increasing the parameter $\beta$. In lower panel, the behavior of
deceleration parameter is studied by fixing $\beta=0.5016$ and
varying $\alpha$. We see that $q$ becomes larger
by increasing $\alpha$. Here, we find the dependency of the deceleration parameter
$q$ on the parameters of new HDE model.\\
Calling Eq. (\ref{solv}), we plot the evolution of dimensionless
Hubble parameter, $E(a)$, for new HDE model in Fig. (3). In upper
panel, we fix $\alpha=0.8824$
 and vary the parameter $\beta$. The smaller
value the parameter  $\beta$ is taken, the bigger the Hubble
parameter expansion rate $E(a)$ can reach. In lower panel, by fixing
$\beta=0.5016$, we vary the parameter $\alpha$. The dimensionless
Hubble parameter $E$ is bigger for larger value of $\alpha$  at any
scale factor $a<1$. While for $a>1$, $E$ is bigger for smaller
 value of $\alpha$. From this figure, we find that both the model parameters $\alpha$ and $\beta$ can impact the cosmic
 expansion history in new HDE model.\\

Finally, we discuss the statefinder diagnostic for new HDE model.
  The statefinder pair \{r,s\} in this model is given by Eqs. (\ref{r2}) and
(\ref{s2}). One can easily see the dependency of \{r,s\} on the
parameters of new HDE model as well as the curvature parameter
$\Omega_k$, in Eqs. (\ref{r2}) and (\ref{s2}). In spatially flat
universe, where $\Omega_k=0.0$, the parameters $r$ and $s$  reduce
as

\begin{eqnarray}\label{r3}
&r=1+\frac{-2\alpha+3\beta+2}{\beta^2}\times  \\ \nonumber
&\Big[(\frac{2}{-2\alpha+3\beta+2}\Omega_{m0}-1)(\alpha-1)e^{-\frac{2(\alpha-1)x}{\beta}}\Big]\\
\nonumber&
\Big{/}\Big[-(\frac{2}{-2\alpha+3\beta+2}\Omega_{m0}-1)(\alpha-1)e^{-\frac{2(\alpha-1)x}{\beta}}\\
\nonumber & +\frac{2}{-2\alpha+3\beta+2}\Omega_{m0}e^{-3x} \Big]
\end{eqnarray}
\begin{eqnarray}\label{s3}
s=\frac{2(\alpha-1)}{3\beta}
\end{eqnarray}
From Eq. (\ref{s3}), we see that the parameter $s$ is independent of
cosmic scale factor in spatially flat universe. By Choosing
$\alpha=1, \beta\neq0$, we get \{r=1,s=0\} which is coincide to the
location of spatially flat $\Lambda$CDM model in statefinder
plane.\\
In Fig.(4), we show the evolutionary trajectories of new HDE model
in statefinder plane for non flat universe with $\Omega_k=0.0305$.
In this diagram, the standard $\Lambda$CDM model in spatially flat
universe corresponds to a fixed point $\{r=1,s=0\}$ indicated by
star symbol. In upper panel, by fixing $\alpha=0.8824$, we choose
the illustrative values $0.4$, $0.5$ and $0.6$ for $\beta$.
 While the universe
expands, the trajectories of the statefinder start from right to
left. The parameter $r$ increases while the parameter $s$ decreases.
The color circles on the curves represent the today's value of the
statefinder parameter ($s_0,r_0$). The current data of $s$ and $r$
are valuable, if they can be extracted from coming data of SNAP
(SuperNova Acceleration Probe) experiments. Hence, the statefinder
diagnostic combined with future SNAP observation can be useful to
discriminate between various dark energy models. Here, we can easily
see that the statefinder trajectories are dependent on the parameter
$\beta$ of new HDE model. Different values of $\beta$ give the
different evolutionary trajectories in $\{r, s\}$ plane. Also,
distance of the point ($s_0,r_0$) to $\Lambda$CDM fixed point
becomes shorter for larger value of $\beta$. We can also see that
for larger value of $\beta$, the present value $s_0$ increases and
the present value $r_0$
decreases.\\
In lower panel, by fixing $\beta=0.5016$, the evolutionary
trajectories in $s-r$ diagram is plotted for different values of the
parameter $\alpha$. Same as upper panel, by expanding the universe,
the trajectories start from right to left. The parameter $r$
increases while the parameter $s$ decreases. Also, it can be seen
that the statefinder trajectories are dependent on the parameter
$\alpha$ of new HDE model. Different values of $\alpha$ give the
different evolutionary trajectories in this plane. The distance of
the point ($s_0,r_0$) to $\Lambda$CDM fixed point becomes shorter
for larger value of $\alpha$. Like the effect of $\beta$, the
present value
$s_0$ increases while the present value $r_0$ decreases, by taking the larger value of $\alpha$.\\
In Fig.(5), by using the best fit values for cosmological
parameters:($\Omega_bh^2=0.0228$, $H_0=0.7020$,
$\Omega_k=0.0305$,$\Omega_{\Lambda 0}=0.6934$,
$\Omega_{m0}=0.2762$), we plot the evolutionary trajectory in $s-r$
plane (upper panel) and $q-r$ plane (lower panel) for fixed
observational values: $\alpha=0.8824$ and $\beta=0.5016$.
 In upper panel the evolutionary
trajectory starts from right at the past time, reaches to
($s_0=-0.13,r_0=1.46$) at the present time (circle point). In lower
panel the evolutionary trajectory in $q-r$ plane starts from
($q=0.5,r=1$) at the past, representing the CDM dominated universe
at the early time, reaches to ($q=-0.55,r=1.46$) at the present
time.

\section{Conclusion}
In this work, we investigated the holographic dark energy model with
new infrared cut-off proposed by Granda and Oliveros in spatially
non-flat universe. Contrary to the HDE model based on event horizon,
this model depends on the local quantities and avoids the causality
problem. Therefore, The new HDE model can be assumed as a
phenomenological model for holographic energy density. Here, we
calculated some relevant cosmological parameters and their
evolution. Also, the statefinder diagnostic is performed for new HDE
model in non-flat universe. In summery,\\
\emph{i)} The EoS parameter, $w_{\Lambda}$, starts from
$w_{\Lambda}>-1$ at the early time and crosses the phantom divide
$w_{\Lambda}<-1$ at the late time. This behavior of $w_{\Lambda}$ is
dependent on the model parameters. The larger value of $\alpha$ and
smaller value of $\beta$ give the larger EoS parameter,
$w_{\Lambda}$. Also for smaller value of $\alpha$ or larger value of
$\beta$, the phantom divide is achieved earlier.\\
In new HDE model, the universe undergoes decelerated expansion at
the early time ($q>0$) and then starts accelerated expansion ($q<0$)
at the later time. The transition epoch from decelerated phase to
accelerated phase occurs sooner by increasing $\beta$ or $\alpha$.\\
 The cosmic expansion history in new HDE model is dependent on
 the model parameters $\alpha$ and
 $\beta$. The smaller value of $\beta$ is taken, the bigger Hubble
parameter can reach. Also, the Hubble parameter become larger by
increasing $\alpha$ at $a<1$ and decreasing $\alpha$ at $a>1$.\\
\emph{ii)}We studied the new HDE model from the viewpoint of
statefinder diagnostic. The statefinder diagnostic is a crucial tool
for discriminating different DE models. Also, the present value of
\{r, s\} can be viewed as a discriminator for testing different DE
models if it can be extracted from precise observational data in a
model-independent way. We calculate the evolution of new HDE model
in the statefinder plane for different values of the model parameter
$\alpha$ and $\beta$. The statefinder trajectories are dependent on
the model parameters . Different values of $\alpha$ and $\beta$ are
taken, different evolutionary trajectories are achieved. By
expanding the universe, the trajectories start from right to left in
$s-r$ plane, the parameter $s$ decreases and $r$ increases. Distance
of the present value ($s_0,r_0$) from the $\Lambda$CDM fixed point
($s=0,r=1$) becomes shorter for larger values of $\beta$ and
$\alpha$.\\
We also performed the statefinder diagnostic in $s-r$ and $q-r$
planes for new HDE model in the light of best fit results of
SNe+BAO+OHD+CMB experiments. These trajectories yield
($s_0=-0.13,r_0=1.46$) and ($q=-0.55,r=1.46$) at present time. The
evolutionary trajectory in $q-r$ plane starts from ($q=1/2,r=1.0$)
which is coincidence on the location of CDM model in $s-r$ plane.\\
Finally, it is of interest to compare the nwe HDE model and
holographic DE (HDE) model from the viewpoint of statefinder
diagnostic. The statefinder diagnostic for HDE model in non-flat
universe is preformed in \citep{r16}. In the light of best fit
result of the SN+CMB data analysis, the evolutionary trajectories in
$s-r$ and $q-r$ planes gives the present
values:($s_0=-0.102,r_0=1.357$) and ($q_0=-0.590,r_0=1.357$) for HDE
model in non-flat universe \citep{r16}. Therefore the distance from
the $\Lambda$CDM fixed point ($s=0,r=1.0$) is  shorter for new HDE
model compare with HDE model. As a similarity, for both HDE and new
HDE models, the trajectories in $q-r$ plane starts from $q=1/2,r=1$
at the early time which is denoting the CDM-dominated universe. WE
hope that the future high-precision SNAP-type observations can
determine the statefinder parameters and consequently single out the
right cosmological DE models.

\noindent{{\bf Acknowledgements}}\\
This work has been supported financially by Research Institute for
Astronomy $\&$ Astrophysics of Maragha (RIAAM), Maragha, Iran.
\newpage

\begin{figure}[!htb]
\includegraphics[width=8cm]{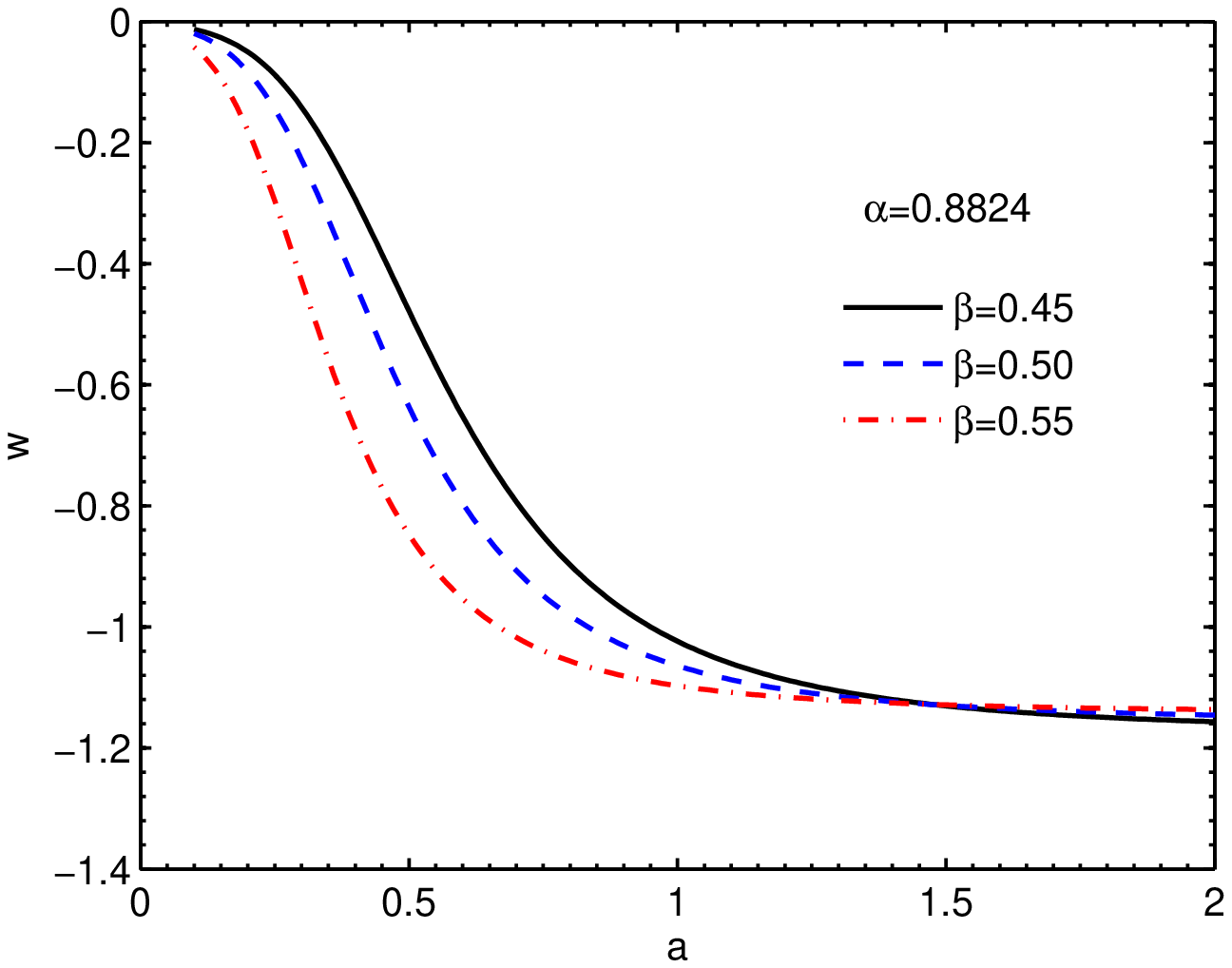} \includegraphics[width=8cm]{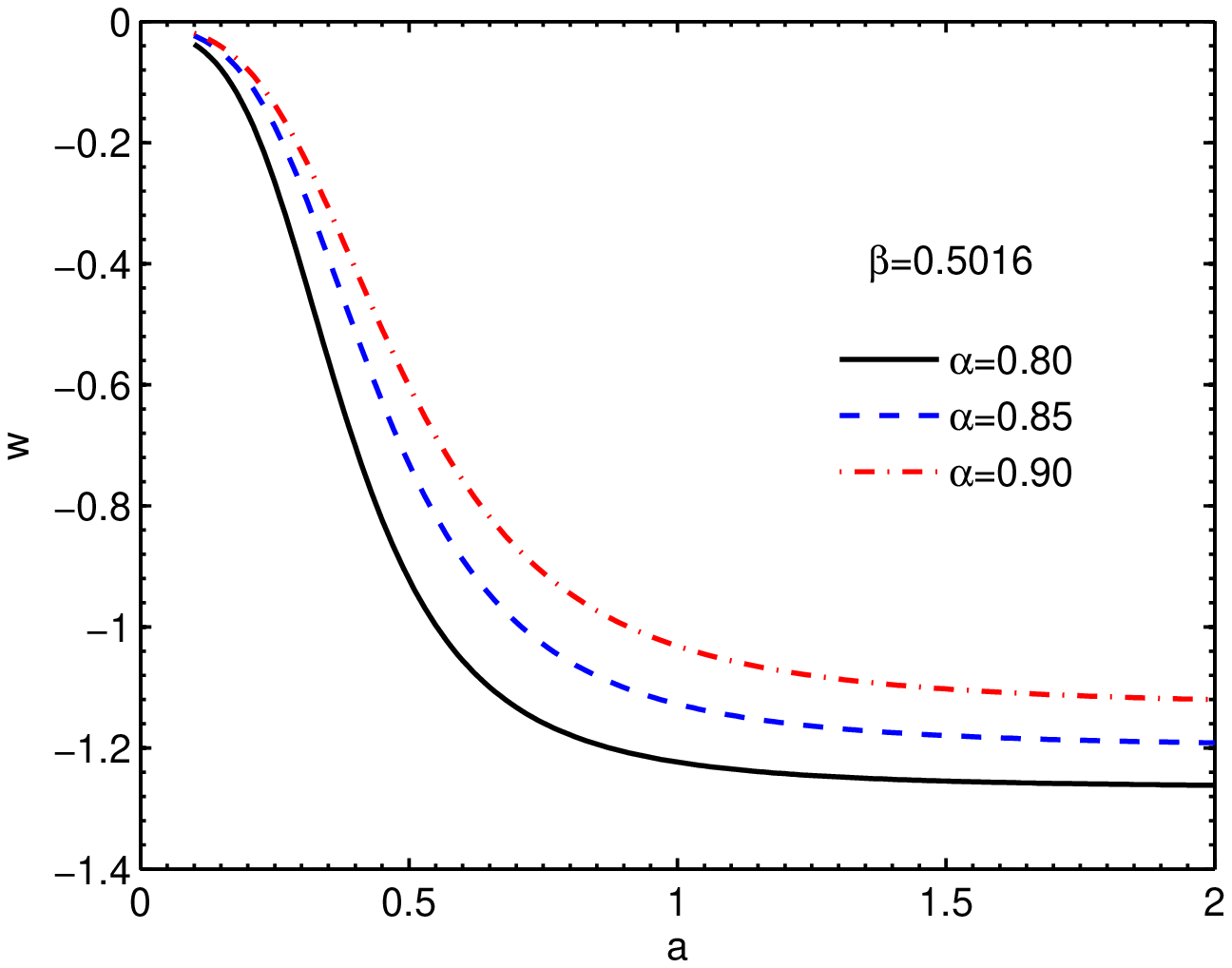} %
\caption{The evolution of EoS parameter of new HDE model,
$w_{\Lambda}$, versus scale factor $a$ for different values of model
parameters $\alpha$ and $\beta$ in non flat universe with
$\Omega_k=0.0305$. In upper panel, by fixing $\alpha$ as a best fit
value: $\alpha=0.8824$, we vary $\beta$ as $0.45, 0.50 , 0.55$
corresponding to black solid line, blue dashed line and red
dotted-dashed line, respectively. In lower panel, by fixing $\beta$
as a best fit value: $\beta=0.5016$, $\alpha$ is varied as $0.80$
(black solid line), $0.85$ (blue dashed line), $0.90$ (red
dotted-dashed line).}
\end{figure}

\begin{figure}[!htb]
\includegraphics[width=8cm]{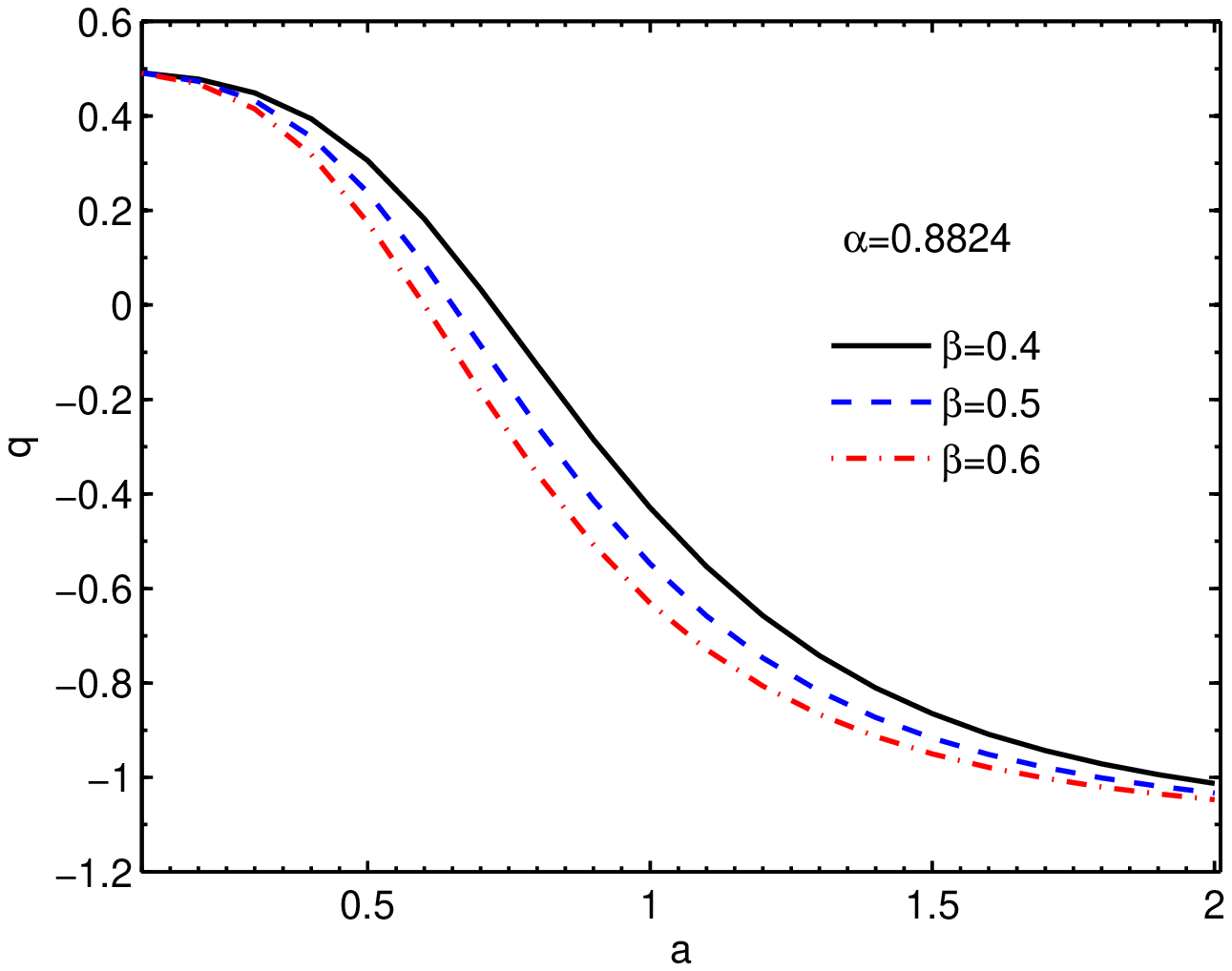} \includegraphics[width=8cm]{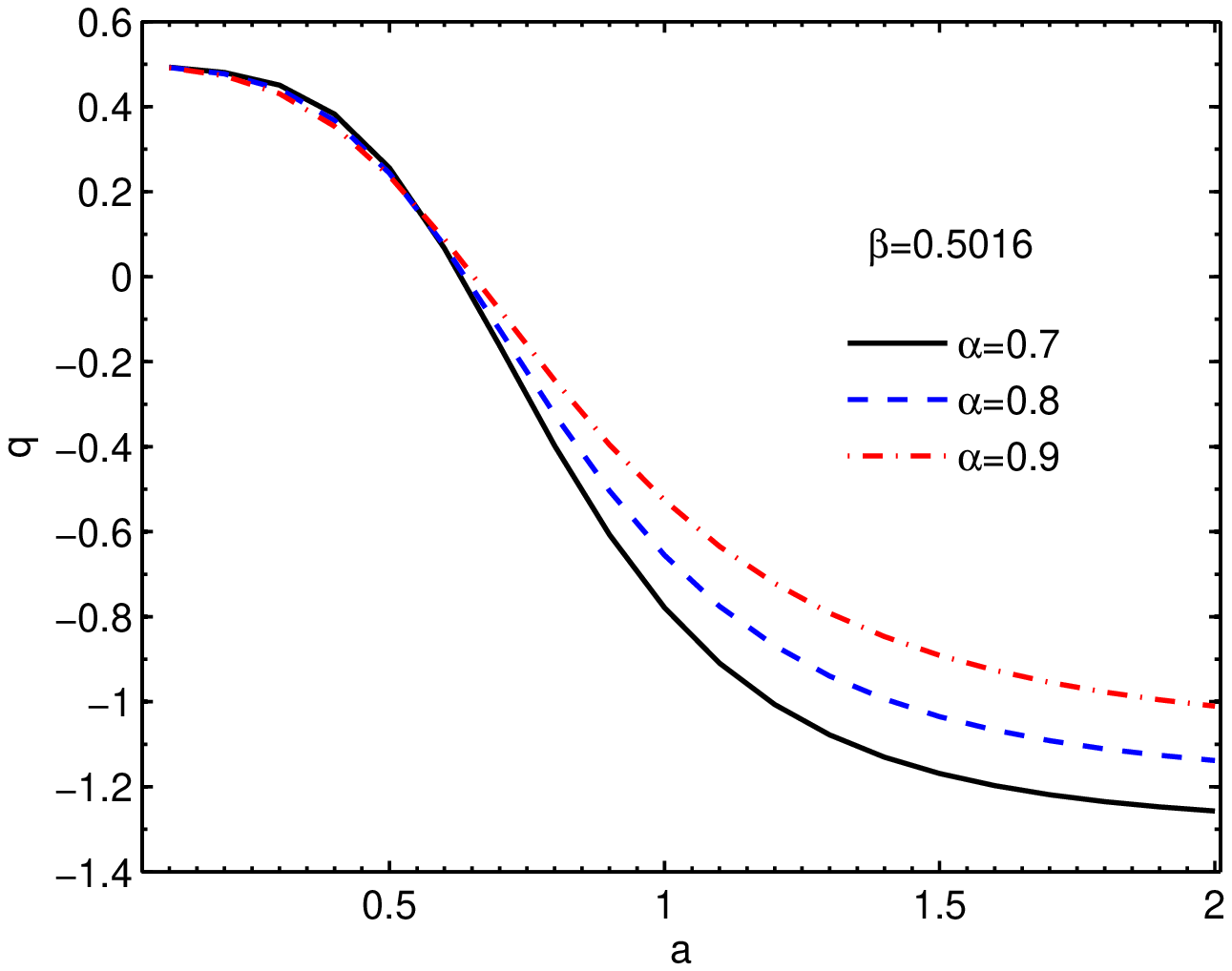} %
\caption{The evolution of deceleration parameter $q$ in new HDE
model versus scale factor $a$ for different  values of model
parameters $\alpha$ and $\beta$ in non flat universe with
$\Omega_k=0.0305$. In upper panel, by fixing $\alpha$, we vary
$\beta$ as $0.4, 0.5 , 0.6$ corresponding to black solid line, blue
dashed line and red dotted-dashed line, respectively. In lower
panel, by fixing $\beta$, $\alpha$ is varied as $0.7$ (black solid
line), $0.8$ (blue dashed line), $0.9$ (red dotted-dashed line).}
\end{figure}

\begin{figure}[!htb]
\includegraphics[width=8cm]{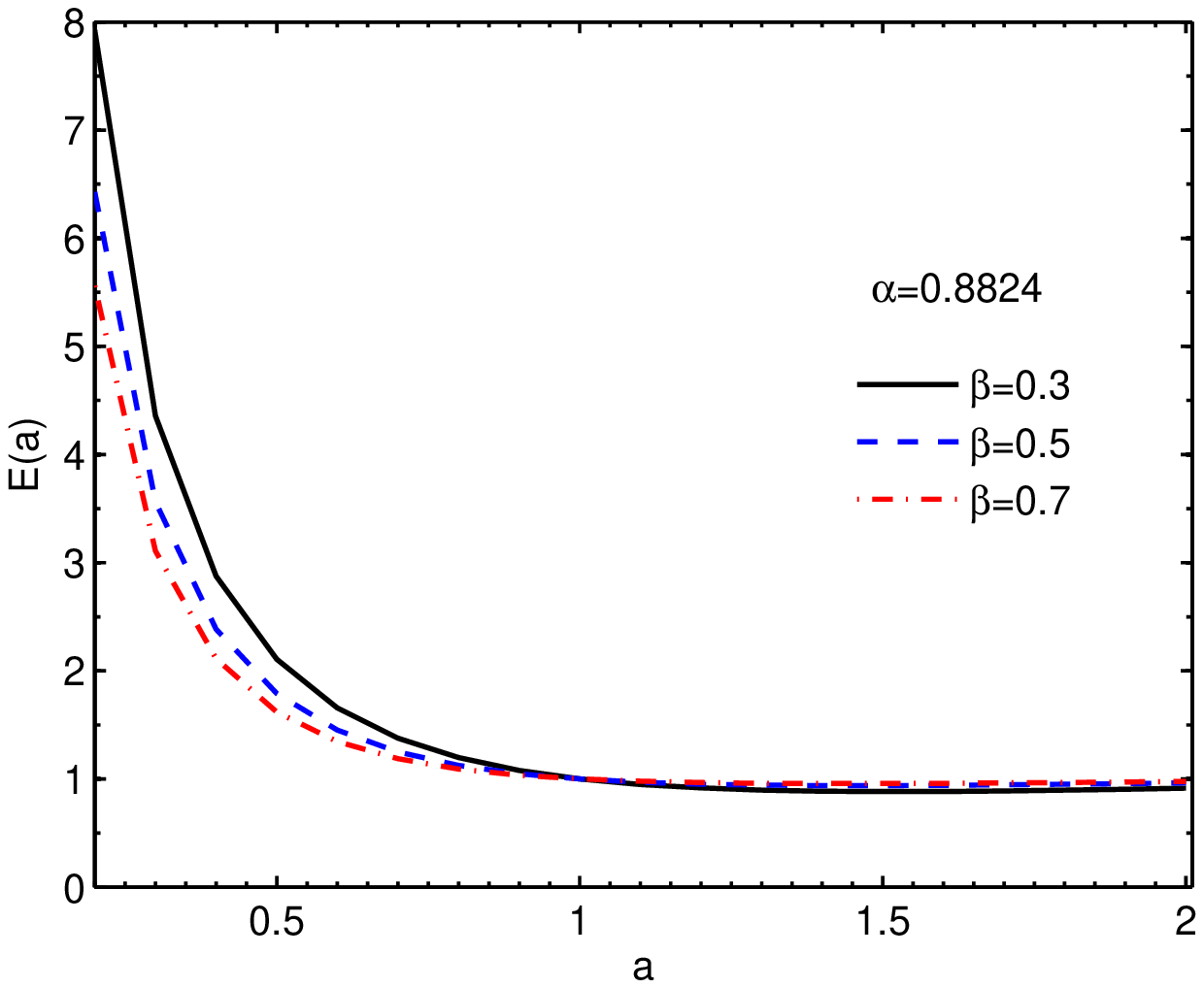} \includegraphics[width=8cm]{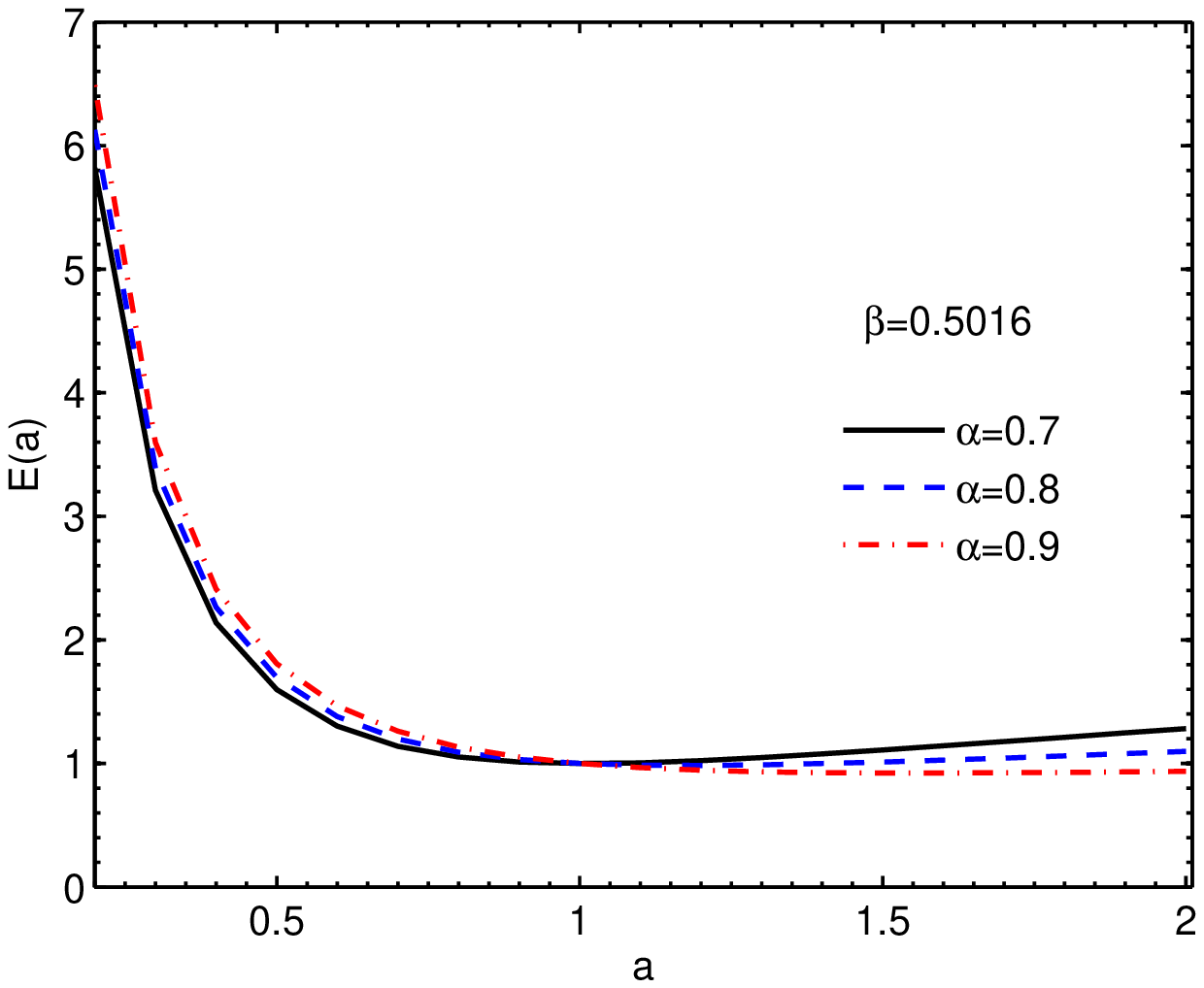} %
\caption{ Cosmological evolution of dimensionless Hubble parameter,
$E$, as a function of  scale factor  in non flat universe for new
HDE model. In upper panel, we choose the observational best fit
values: $\alpha=0.8824$ and $\Omega_k=0.0305$ and vary the parameter
$\beta$ as $0.3$, $0.5$ and $0.7$ corresponding to black solid, blue
dashed and red dotted-dashed lines, respectively.  In lower panel,
by fixing $\beta=0.5016$, $\alpha$ is varied as $0.7$, $0.8$, $0.9$
corresponding to black solid line, blue dashed  line and  red
dotted-dashed line,
 respectively.}
\end{figure}

\begin{figure}[!htb]
\includegraphics[width=8cm]{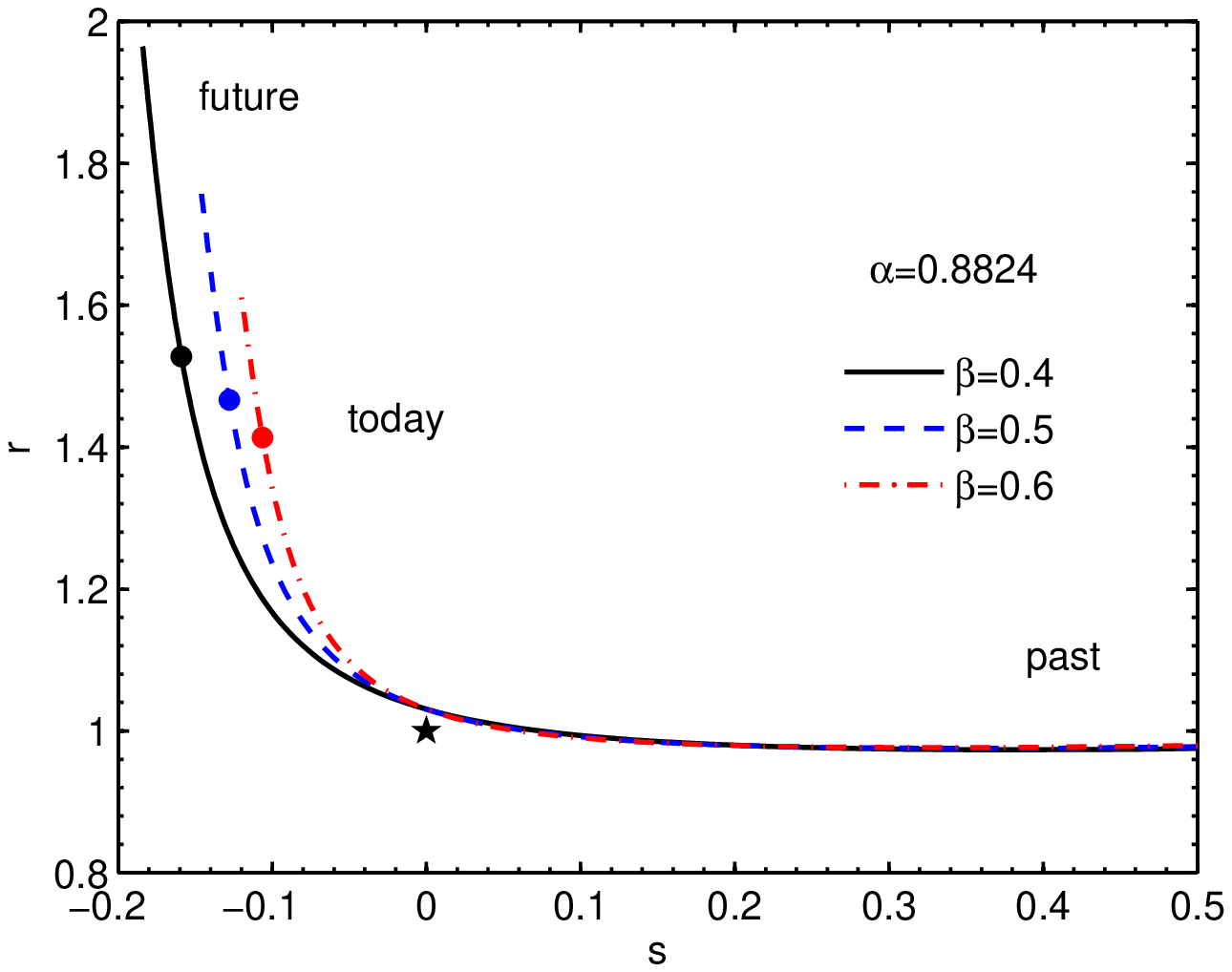} \includegraphics[width=8cm]{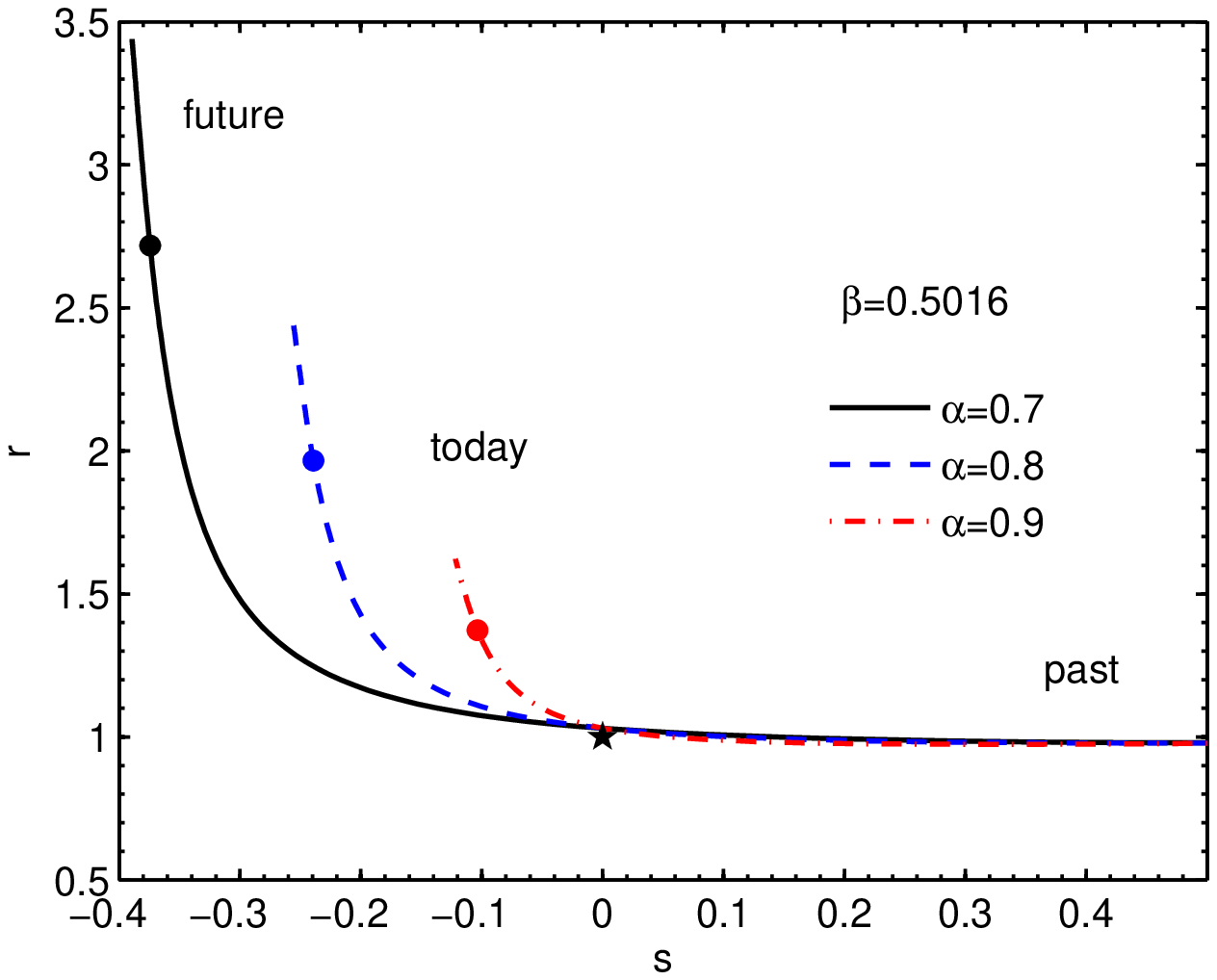} %
\caption{An illustrative example for the statefinder diagnostic of
new HDE model in non flat universe. In upper panel, the evolutionary
trajectories in $s-r$ plane are plotted, by fixing $\alpha=0.8824$
and varying $\beta$ as $0.4$, $0.5$ and $0.6$ corresponding to black
solid line, blue dashed line and red dotted dashed line,
respectively. The circle point on the curves show the today's value
of statefinder parameters ($s_0,r_0$). The star symbol indicates the
location of standard flat $\Lambda$CDM model in $s-r$
plane:$\{s=0,r=1\}$. In lower panel, the evolutionary trajectories
are plotted for different illustrative values of $\alpha$, by fixing
$\beta=0.5016$. The evolutionary trajectories of illustrative cases
$\alpha=0.7$, $\alpha=0.8$ and $\alpha=0.9$ have been shown by black
solid line, blue dashed line and red dotted-dashed line,
respectively. Circle point on the curves denotes the today's value
($s_0,r_0$) in $s-r$ plane. Same as upper panel, The star symbol
indicates the location of standard flat $\Lambda$CDM model. }
\end{figure}

\begin{figure}[!htb]
\includegraphics[width=8cm]{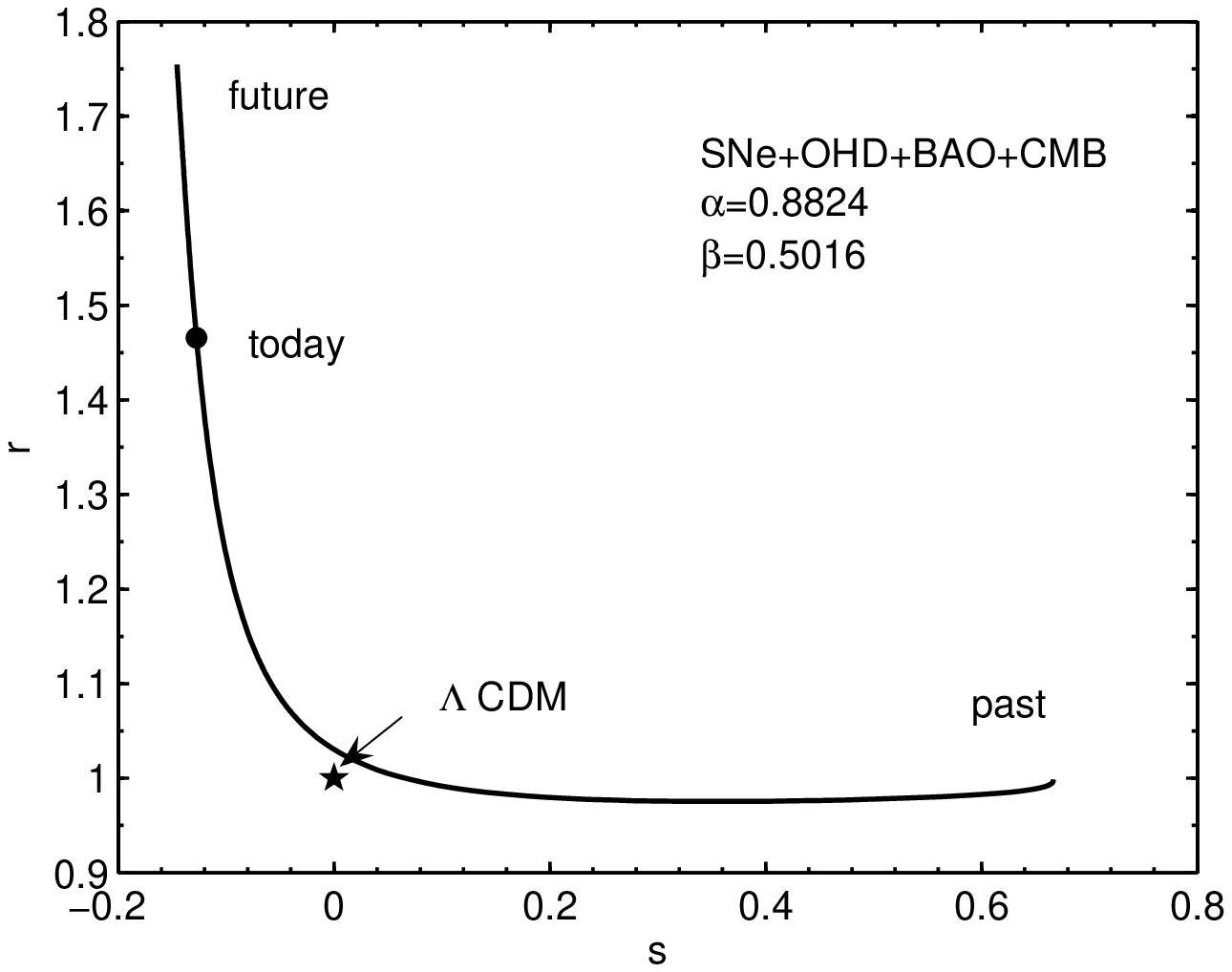} \includegraphics[width=8cm]{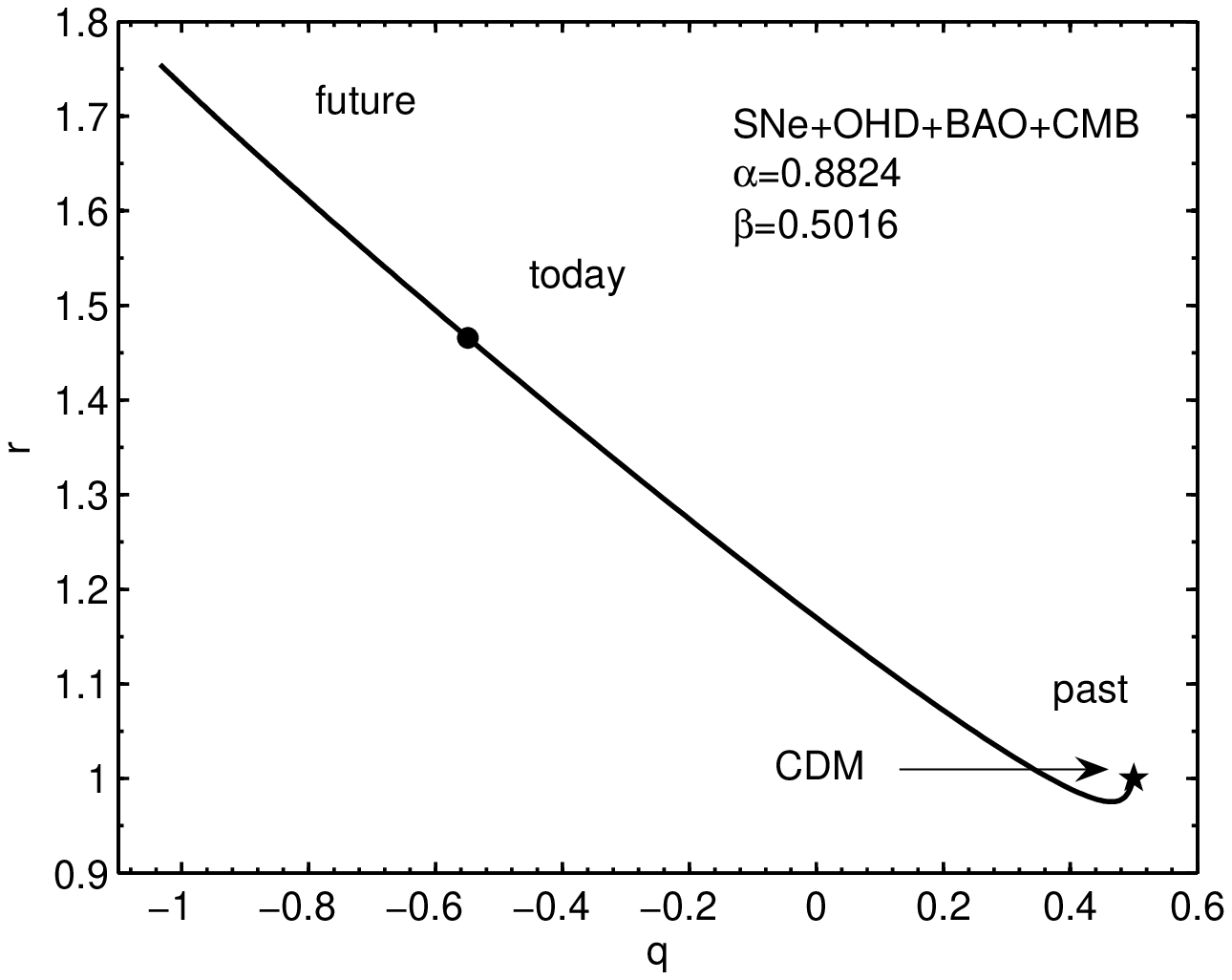} %
\caption{ The statefinder diagrams $r(s)$ (upper panel) and $r(q)$
(lower panel) for new HDE model in the light of best fit results of
SNe+OHD+BAO+CMB experiments. The circles on the curves indicates the
present value ($s_0,r_0$)in upper panel, and ($q_0,r_0$)in lower
panel. The star symbol on the upper panel indicates the location of
standard $\Lambda CDM$ and in the lower panel represents the CDM
dominated universe. }
\end{figure}



\newpage
\begin{thebibliography}{99}

\bibitem[Alam et al. (2003)]{alam} Alam, U.,~Sahni, V.,~Saini, T.~D.,~\&
Starobinsky, A.~A.,~MNRAS. \textbf{344}, 1057 (2003).

\bibitem[Alam et al. (2004)]{alam3}
Alam, U., Sahni, V., Starobinsky, A.A., J. Cosmol. Astropart. Phys.
06, 008 (2004); Huterer, D., \& Cooray, A., Phys. Rev. D 71, 023506
(2005); Wang, Y., \& Tegmark, M., Phys. Rev. D 71, 103513 (2005).

\bibitem[Allen et al. (2008)]{ref:07060033}
 Allen, S. W., Rapetti, D. A., Schmidt, R. W., Ebeling, H., Morris, R. G., \& Fabian, A.
 C., Mon. Not. Roy. Astron. Soc. {\bf 383} 879 (2008).

\bibitem[Amendola (2000)]{intde}
  Amendola, L.,~Phys. Rev. D {\bf 62}, 043511 (2000);
  Comelli, D.,~Pietroni, M.,~\& Riotto, A.,~Phys. Lett. B {\bf 571}, 115 (2003);
   Setare, M.R., JCAP \textbf{0701}, 023, (2007); Jamil, M., \& Rashid, M.A., Eur. Phys. J. C \textbf{60} (2009) 141; ibid,
Eur. Phys. J. C \textbf{58}, 111,2008; Setare, M.~R.,~\& Saridakis,
E.~N.,~Phys. Lett. B {\bf 670}, 1 (2008).

\bibitem[Bennett et al. (2003)]{Bennett} Bennett, C.L., et al., Astrophys. J. Suppl. {\bf 148}, 1 (2003); Spergel,
D.N., Astrophys. J. Suppl. {\bf 148}, 175 (2003); Tegmark M., et
al., Phys. Rev. D {\bf 69}, 103501 (2004); Seljak, U., Slosar, A.,
\& McDonald, P., J. Cosmol. Astropart. Phys. {\bf 10}, 014 (2006);
Spergel, D.N., et al., Astrophys. J. Suppl. {\bf 170}, 377 (2007).

\bibitem[Cai (2007)]{age1} Cai, R. G., Phys. Lett. B {\bf 657} 228 (2007);
Wei, H., \& Cai, R. G., Phys. Lett. B {\bf 660} 113(2008).

\bibitem[Caldwell (2002)]{Caldwell1} Caldwell, R.R., Phys. Lett. B {\bf 545}, 23
(2002); Nojiri, S., \& Odintsov, S.D., Phys. Lett. B {\bf 562}, 147
(2003); Nojiri, S., \& Odintsov, S.D., Phys. Lett. B {\bf 565}, 1
(2003).

\bibitem[Caldwell \& Kamionkowski (2004)]{Caldwell} Caldwell, R.R., \& Kamionkowski, M., [astro-ph/0403003];
Wang, B., Gong, Y. G., \& Su, R. K., Phys. Lett. B 605, 9 (2005).

\bibitem[Chang (2006)]{obs2} Chang, Z., Wu ,~F.~Q., \& Zhang, X.,~Phys. Lett. B {\bf 633}, 14 (2006).

\bibitem[Chang et al. (2007)]{r18} Chang, B.~R.,~Liu, H.~Y.,~Xu, L.~X.,~Zhang C.~W.,~\& Ping, Y.~L.,~
JCAP \textbf{0701}, 016 (2007).

\bibitem[Chiba et al. (2000)]{Chiba} Chiba, T., Okabe, T., \& Yamaguchi, M., Phys. Rev. D {\bf 62},
023511 (2000); Armend\'{a}riz-Pic\'{o}n, C., Mukhanov, V., \&
Steinhardt, P.J., Phys. Rev. Lett. {\bf 85}, 4438 (2000);
Armend\'{a}riz-Pic\'{o}n, C., Mukhanov, V., \& Steinhardt, P.J.,
Phys. Rev. D {\bf 63}, 103510 (2001).

\bibitem[Cohen et al. (1999)]{Cohen} Cohen, A.~G., Kaplan, D. B., \& Nelson, A.~E., Phys. Rev. Lett. {\bf 82}, 4971 (1999).

\bibitem[Copeland et al. (2006)]{copel}
Copeland, E. J. , Sami, M. \& Tsujikawa, S., Int. J. Mod. Phys. D
\textbf{15}, 1753(2006).

\bibitem[Deffayet et al. (2002)]{Deffayet} Deffayet, C., Dvali, G.R., \& Gabadadze, G., Phys.
Rev. D {\bf 65}, 044023 (2002); Sahni, V., \& Shtanov, Y., J.
Cosmol. Astropart. Phys. {\bf 11}, 014 (2003).

\bibitem[Eisenstein et al. (2005)]{SDSS}
 Eisenstein, D.J., et al., Astrophys. J. {\bf 633}, 560 (2005).

\bibitem[Elizalde et al. (2004)]{Elizalde1} Elizalde, E., Nojiri, S., \& Odinstov, S.D., Phys.
Rev. D {\bf 70}, 043539 (2004); Nojiri, S., Odintsov, S.D., \&
Tsujikawa, S., Phys. Rev. D {\bf 71}, 063004 (2005); Anisimov, A.,
Babichev, E., \& Vikman, A., J. Cosmol. Astropart. Phys. {\bf 06},
006 (2005).

\bibitem[Enqvist et al. (2005)]{obs3}
  Enqvist, K.,Hannestad,~S., \& Sloth, M.~S.,~JCAP {\bf 0502} 004 (2005);
  Shen, J.,~Wang, B.,~Abdalla, E.,~\& Su, R.~K.,~Phys. Lett. B {\bf 609} 200 (2005);
  Kao, H.~C.,~Lee, W.~L.,~\& Lin, F.~L.,~Phys. Rev. D {\bf 71} 123518 (2005).


\bibitem[Frieman et al. (2008)]{fri} Frieman, J.A., Turner, M.S., \& Huterer,
D., [arXiv:0803.0982].

\bibitem[Gasperini et al. (2002)]{Gasperini} Gasperini, M., Piazza, F., \& Veneziano, G., Phys. Rev. D {\bf 65},
023508 (2002); Arkani-Hamed, N., Creminelli, P., Mukohyama, S., \&
Zaldarriaga, M., J. Cosmol. Astropart. Phys. {\bf 04}, 001 (2004);
Piazza, F., \& Tsujikawa, S., J. Cosmol. Astropart. Phys. {\bf 07},
004 (2004).

\bibitem[Granda \& Oliveros (2008)]{garanda} Granda, L.N., \& Oliveros, A., Phys. Lett. B {\bf 669}, 275
(2008).

\bibitem[Granda \& Oliveros (2009)]{granda2} Granda, L.N., \& Oliveros, A., Phys. Lett. B {\bf 671}, 202 (2009);
Karami, K., \& Fehri, J., Phys. Lett. B {\bf 684}, 61, (2010).

\bibitem[Hicken et al. (2009)]{397Constitution}
  Hicken M.,~et al.,  Astrophys. J. {\bf 700} 1097 (2009).

\bibitem[Hsu (2004)]{holo1} Hsu, S. D. H., Phys. Lett. B {\bf 594} 13 (2004);
 Li, M., Phys. Lett. B {\bf 603} 1 (2004).

\bibitem[Huang \&~Li (2004)]{nonflat} Huang, Q.~G.,~\& Li, M.,~JCAP {\bf 0408}, 013 (2004).

\bibitem[Huang \& Gong (2004)]{obs3a} Huang, Q.~G., \& Gong, Y.~G.,~JCAP {\bf 0408}, 006 (2004).

\bibitem[Ito (2005)]{holoext}
 Ito, M.,~Europhys.\ Lett.\  {\bf 71}, 712 (2005);
Enqvist K.,~\& Sloth, M.~S.,~Phys.\ Rev.\ Lett.\  {\bf 93}, 221302
(2004); Huang, Q.~G.~\& Li, M.,~JCAP {\bf 0503}, 001 (2005);
  Pavon, D.~\& Zimdahl, W.,~Phys. Lett. B {\bf 628}, 206 (2005);
Wang, B.,~Gong, Y.,~\& Abdalla, E.,~Phys. Lett. B {\bf 624}, 141
(2005); Kim, H.,~Lee, H.~W.,~\& Myung, Y.~S.,~Phys. Lett. B {\bf
632}, 605 (2006); Nojiri, S.,~\& Odintsov, S.~D.,~Gen. Rel. Grav.
{\bf 38}, 1285 (2006); Elizalde, E.,Nojiri, S., Odintsov, S.~D., \&
Wang, P.,~Phys. Rev. D {\bf 71}, 103504 (2005); Hu B.,~\& ~Ling, Y.,
Phys. Rev. D {\bf 73}, 123510 (2006); Li, H.,~Guo, Z.~K.,~\& Zhang,
Y.~Z.,~Int. J. Mod. Phys. D {\bf 15}, 869 (2006); Setare,
M.~R.,~Phys. Lett. B {\bf 642}, 1 (2006); Setare, M.~R.,~Phys. Lett.
{\bf B642}, 421, (2006); Saridakis, E.~N.,~Phys. Lett. B {\bf 660},
138 (2008); Saridakis, E.~N.,~JCAP {\bf 0804}, 020 (2008);
Saridakis, E.~N.,~Phys. Lett. B {\bf 661}, 335 (2008); \emph{M. R.
Setare, Phys. Lett.B \textbf{644}, 99 (2007); M. R. Setare, Eur.
Phys. J. C \textbf{50}, 991 (2007); M. R. Setare, Phys. Lett. B
\textbf{648}, 329 (2007); M. R. Setare, Phys. Lett. B \textbf{653},
116 (2007);}.

\bibitem[Kamenshchik et al. (2001)]{Kamenshchik} Kamenshchik, A., Moschella, U., \& Pasquier, V., Phys. Lett. B {\bf 511}, 265
(2001); Bento, M.C., Bertolami, O., \& Sen, A.A., Phys. Rev. D {\bf
66}, 043507 (2002).

\bibitem[Kerner (1982)]{kerner}
 Kerner, R., Gen. Rel. Gravit. \textbf{14}, 453 (1982); Duruisseau, J. P., \& Kerner, R., Class. Quantum
Grav. \textbf{3}, 817 (1986).

\bibitem[Khodam-Mohammadi \& Malekjani (2010)]{khod10}
Khodam-Mohammadi, A., \& Malekjani, M., Astrophys. Space Sci.
DOI(10.1007/s 10509-010-0422-y) (2010), [gr-qc/1003.0543].

\bibitem[Komatsu et al. (2009)]{5yWMAP}
 Komatsu, E.,~et al., Astrophys. J. Suppl. {\bf 180} 330 (2009);
 Dunkley, J., et al., [arXiv:0803.0586].

\bibitem[Li (2004)]{Li} Li, M.,~Phys. Lett. B {\bf 603}, 1 (2004).

\bibitem[Malekjani \& Khodam-Mohammadi (2010)]{malek} Malekjani, M., \& Khodam-Mohammadi, A., Int. J. Mod.
Phys. D, {\bf 19}, 1 (2010). [arXiv:1004.0508].

\bibitem[Percival et al. (2009)]{ref:Percival2} Percival, W.J., et al., [arXiv:0907.1660].

\bibitem[Perlmutter et al. (1998)]{SN} Perlmutter, S. et al., Nature \textbf{391}, 51
(1998); Riess, A.G. et al., Astron. J. \textbf{116}, 1009 (1998);
Riess, A.G. et al., Astron. J. \textbf{117},707 (1999); Spergel, D.
N. et al. [WMAP Collaboration], Astrophys. J. Suppl. 148, 175 (2003)
[astro-ph/0302209]; Spergel, D. N. et al., astro-ph/0603449;
 Tegmark, M. et al. [SDSS Collaboration], Phys. Rev. D
69, 103501 (2004) [astro-ph/0310723];
 Abazajian, K. et al. [SDSS Collaboration], Astron. J. 128, 502 (2004)
[astro-ph/0403325]; Abazajian, K. et al. [SDSS Collaboration],
Astron. J. 129, 1755 (2005) [astro-ph/0410239].

\bibitem[Sahni \& Starobinsky (2000)]{sahnicc}
 Sahni, V., \& Starobinsky, A. A., Int. J. Mod. Phys. D \textbf{9}, 373 (2000), [astro-ph/9904398];
S. M.  Living Rev. Rel. \textbf{4}, 1 (2001), [astro-ph/0004075];
Carroll, P., Peebles, J. E., \& Ratra, B., Rev. Mod. Phys.
\textbf{75}, 559 (2003), [astro-ph/0207347]; Padmanabhan, T., Phys.
Rept. \textbf{380}, 235 (2003), [hep-th/0212290]; Copeland, E. J.,
Sami, M., \& Tsujikawa, S., Int. J. Mod. Phys. D \textbf{15}, 1753
(2006), [hep-th/0603057]; Bousso, R., Gen. Rel. Grav. \textbf{40},
607 (2008).

\bibitem[Sahni et al. (2003)]{Sahni} Sahni, V.,~Saini, T.~D.,~Starobinsky, A.~A.,~\& Alam, U.,~JETP
Lett. \textbf{77}, 201 (2003).


\bibitem[Sen (1999)]{Sen} Sen, A., J. High Energy Phys. {\bf 10}, 008 (1999);
Bergshoeff, E.A., de Roo, M., de Wit, T.C., Eyras, E., \& Panda, S.,
J. High Energy Phys. {\bf 05}, 009 (2000); Sen, A., J. High Energy
Phys. {\bf 04}, 048 (2002); Sen, A., J. High Energy Phys. {\bf 07},
065 (2002); Padmanabhan, T., Phys. Rev. D {\bf 66}, 021301 (2002);
Padmanabhan, T., \& Choudhury, T.R., Phys. Rev. D {\bf 66}, 081301
(2002); Abramo, L.R.W., \& Finelli, F., Phys. Lett. B {\bf 575}, 165
(2003).

\bibitem[Setare et al. (2007)]{r16} Setare, M.~R.,~Zhang, J.,~\& Zhang, X.,~JCAP \textbf{0703}, 007
(2007).

\bibitem[Setare (2010)]{setare}
 Setare, M. R., Astrophys. Space Sci. {\bf 326}, 27, 2010;
 Khodam-Mohammadi, A., Majari, P. \& Malekjani, M., Astrophys.Space Sci. DOI(10.1007/s 10509-010-0480-1) [arXiv: 1007. 2705].

\bibitem[Shao \& Gui (2007)]{r22} Shao, Y.,~\& Gui, Y.,~[gr-qc/0703111].

\bibitem[Sievers et al. (2003)]{Sie} Sievers, J. L., et al., Astrophys. J. \textbf{591}, 599 (2003);
Netterfield, C.B., et al., Astrophys. J. \textbf{571}, 604 (2002);
Benoit, A., et al., Astron. Astrophys. \textbf{399} (2003) L25;
Benoit, A., et al., Astron. Astrophys.\textbf{399} (2003) L19.

\bibitem[Simon et al. (2005)]{OHD}
 Simon, J., Verde, L., \& Jimenez, R., Phys. Rev. D {\bf 71} 123001 (2005).

\bibitem[Starobinsky (1980)]{star}
 Starobinsky, A. A., Phys. Lett. B {\bf 91}, 99 (1980).

\bibitem['t Hooft (1993)]{Horava}
  't Hooft, G., [gr-qc/9310026];
  Susskind, L., J. Math. Phys. {\bf 36}, 6377 (1995);
  Horava, P., \& Minic, D., Phys. Rev. Lett. {\bf 85}, 1610 (2000);
  Thomas, S.~D.,~Phys. Rev. Lett. {\bf 89}, 081301 (2002).

\bibitem[Wang et al. (2006)]{wang22}
Wang, B., Gong, Y., Abdalla, E., Phys. Rev. D 74, 083520 (2006).

\bibitem[Wang \& Xu (2010)]{wang2}
Wang, Y., \& Xu, L., Phys. Rev. D {\bf 81}, 083523 (2010).

\bibitem[Wei \& Cai (2007)]{wei} Wei, H., Cai, R. G., Phys. Lett. B \textbf{655}, 1(2007).

\bibitem[Wei \& Zhang (2007)]{zhang4}
Wei, H., \& Zhang, S. N., [arXiv:0707.2129].


\bibitem[Weinberg (1989)]{berg89} Weinberg, S., Rev. Mod. Phys. \textbf{61}, 1
(1989).

\bibitem[Wetterich (1988)]{Wetterich} Wetterich, C., Nucl. Phys. B {\bf 302}, 668
(1988); Ratra, B., \& Peebles, J. Phys. Rev. D {\bf 37}, 321 (1988).

\bibitem[Wu et al. (2008)]{Wu:2007fs}
  Wu, Q., Gong, Y., Wang,~A.,~\& Alcaniz, J.~S.,~Phys. Lett. B {\bf 659}, 34 (2008);
  Ma, Y.~Z., \& Gong, Y.,~Eur. Phys. J. C {\bf 60}, 303 (2009).

\bibitem[Yin-Zhe Ma (2008)]{observ}
Yin-Zhe Ma, Nuc. Phys. B \textbf{804}, 262285 (2008); Daly, R. A. et
al., [arXiv:0710.5345].

\bibitem[Zhang \& Wu (2005)]{obs1}
  Zhang X.,~\& Wu, F.~Q.,~Phys. Rev. D {\bf 72}, 043524 (2005).

\bibitem[Zhang (2005a]{r13} Zhang, X.,~Phys.\ Lett.\ B \textbf{611}, 1 (2005).

\bibitem[Zhang (2005b)]{r14} Zhang, X.,~Int.\ J.\ Mod.\ Phys.\ D \textbf{14}, 1597
(2005).

\bibitem[Zhang et al. (2007)]{r15} Zhang, J.,~Zhang, X.,~\& Liu, H.,~[arXiv:0705.4145].

\bibitem[Zhang et al. (2010)]{zhang} Zhang, L., Cui, J., Zhang, J., \& Zhang, X., Int. J. Mod. Phys. D
\textbf{19}, 21 (2010).

\bibitem[Zimdahl \& Pavon (2004)]{r12} Zimdahl, W.,~\& Pavon, D.,~Gen. Rel. Grav. \textbf{36}, 1483
(2004).

\end{thebibliography}
\end{document}